# Two Kinds of Nothing: What Insignificant Results in Finance Actually Show

David Tan
College of Business Administration, American University of the Middle East, KUWAIT
Department of Applied Finance, Macquarie University, AUSTRALIA

September 2026

Email: d.tan@mq.edu.au

**ABSTRACT**

Claims of the form “we find no evidence that X affects Y” appear throughout the applied finance literature, yet whether such a claim contains evidence of absence or absence of evidence depends entirely on its confidence interval. Still, these intervals are rarely examined by authors – particularly for insignificant results. The term “statistically insignificant” is routinely read to mean zero economic effect. However, a more honest description is that zero could not be rejected along with a range of other coefficient effect sizes. The crucial question is whether effect sizes in that range are consequential. This note distinguishes two kinds of insignificant results that are indistinguishable in a standard regression table: bounded null claims where the intervals reject effect sizes of consequence and thus represent a genuine finding, and vacuous null claims where even consequential effects – along with zero – remain unrejected and therefore establish nothing. The prevailing convention already takes a stand on the threshold for which effects matter: silently and arbitrarily, at wherever the interval's edges happen to fall. I propose a minimal reporting standard for regression results in applied finance, where the smallest consequential effect size is stated (in the units of the decision, per a named increment of the regressor) alongside the descriptive statistics and compared with the relevant edges of the confidence intervals of null claims. Using only the reported coefficient and standard error – and the proposed protocol – authors can distinguish bounded (informative) null claims that reject consequential effect sizes from vacuous null claims that establish no information, perhaps due to deficiencies in data or the identification strategy. The symmetric phrase “no effect” conceals, in particular, the frequent split verdict – bounded in one direction, vacuous in the other. In ongoing work, I apply this framework to published null claims in leading finance journals, beginning with my own.

## 1. The asymmetry

Empirical finance closely polices only one of the two failure modes in inferential statistics: the false positive conclusion. A substantial body of work has examined the heavy reliance and mechanical application of the conventional p-value thresholds and has found that this overstates the evidence for effects. For example, Kim and Ji (2015) conduct a survey that finds leading finance papers applied statistical significance tests mechanically, with little attention paid to the sample size and power of the test[1]. Harvey (2017) argues that the static use of the 5 percent significance threshold is untenable given the scale of multiple testing. In fact, Foster, Smith and Whaley (1997) addressed an adjacent issue two decades earlier, deriving a distribution of maximal R-Squared that accounts for the specification search of the researcher. Brodeur, Lé, Sangnier and Zylberberg (2016) and Brodeur, Cook and Heyes (2020) document the clustering of published test statistics just beyond significance thresholds, indicative of selective reporting and specification searches. And finally, Mitton (2024) shows that reported effect magnitudes in corporate finance are unstandardized, unbenchmarked, and frequently overstated. The cumulative result suggests that statistically significant findings are often less informative than they appear.

The unpoliced failure is the false – or empty – negative. Null claims such as "we find no statistical evidence that X affects Y" are ubiquitous in the literature and rarely accompanied by further investigation. However, the informativeness of null claims is rarely systematically audited in empirical finance. Surveys in the significance-testing literature find that the overwhelming majority of empirical papers dispose of economic magnitude in a sentence or two (McCloskey and Ziliak, 1996; Ziliak and McCloskey, 2008; Kim and Ji, 2015). This inattention applies particularly to insignificant results, where researchers report the coefficient effect as zero and move on. For a literature that painstakingly interrogates whether a significant coefficient is real, there is, in stark contrast, no habit of questioning the informativeness of a null result. Recently, Fitzgerald (2025) conducted an audit of 135 economics papers in leading journals that reported null claims. He found that approximately one to two-thirds of his sample could not rule out economically significant effects. To date, no such comparable audit exists for finance.

[1] The criticism is not that test statistics fail to adjust for sample size – degrees-of-freedom corrections ensure correct Type I error at any n – but that the significance level itself is held fixed while power varies with n, leaving the balance between false positives and missed effects unchosen. The small-sample half of this pathology, where a fixed 5 percent threshold pairs with low power, is the machine that manufactures vacuous nulls.

A statistically insignificant coefficient – by itself – tells us that the null hypothesis of a zero population coefficient cannot be rejected, and nothing else. Yet the evidence behind the null claim that "X has no effect on Y" depends largely on what other population values the data also failed to reject. If the confidence interval is tightly bounded around zero, the data has rejected every effect large enough to matter. I use the term *bounded null* for this case[2]: a genuine finding – that whatever the effect is, it is too small to be of economic consequence – which the applied microeconomics literature calls a "precisely estimated zero". If the interval is wide, economically meaningful effects survive alongside zero; I use the term *vacuous null* for this case. Here the data cannot distinguish consequential effect sizes from zero: the result reports an absence of evidence, not evidence of absence. Bounded and vacuous nulls carry almost opposite meanings, yet a standard reading of a regression table registers only whether zero is rejected and cannot tell the two apart.

This research note is structured as follows: Section 2 develops the bounded/vacuous distinction for null claims – viewing the confidence interval as a set of unrejected hypotheses and setting a three-question protocol for classifying any null claim. Section 3 operationalizes the protocol: the author states the smallest consequential effect size, in decision units per named increment, alongside the descriptive statistics. This facilitates the systematic classification of null claims. A stylized example of classifying null claims drawn from published practice is demonstrated in Section 4. Section 5 draws implications for authors, referees, and readers of existing literature.

The ideas in this note are not new. Neyman (1937) defined the construction of confidence intervals as unrejected sets of hypotheses; a substantial amount of work – some decades old – has been done in emphasizing economic over statistical significance (McCloskey and Ziliak, 1996; Ziliak and McCloskey, 2008; Wasserstein and Lazar, 2016; Amrhein, Greenland and McShane, 2019; Lakens, 2017); equivalence testing provides the formal mechanisms to assess whether effect sizes are indeed negligible (Lakens, 2017); Kim and Robinson (2019) introduced interval-based testing for finance and economics; and Fitzgerald (2025) conducted an audit of 135 null claims across 135 papers in leading economics journals against equivalence benchmarks.

[2] The term "bounded null hypothesis" carries a distinct technical meaning in the randomisation-inference literature (Caughey, Dafoe, Li and Miratrix, 2023), where it denotes a null under which all unit-level treatment effects are weakly negative (or positive). The usage here is unrelated.

This research note contributes to the existing literature as follows: First, I develop a reporting taxonomy and reading rule for applied finance, distinguishing bounded from vacuous null claims directly from standard regression tables without the need for further data analysis. Second, I showcase the common directional cases where a "no effect" claim is bounded on one side and vacuous on the other. Third, I demonstrate that the current convention in blanket "no effect" claims implies a threshold for material effects – albeit silent and arbitrary and typically unknown to the author – as it happens to be where the interval edges happen to fall. Finally, I propose a minimal reporting standard for what constitutes a material effect size, stated in the units of the economic decision per a named increment of the regressor. This makes the economic threshold judgement explicit and contestable.

## 2. Two kinds of nothing

A 95 percent confidence interval is not simply a decorative error band around a point estimate. It is, by construction, a range of null hypotheses that could not be rejected at the 5 percent significance level. The interval is obtained by inverting the hypothesis tests of $H_0$: $\beta = b$ over all candidate values b and reporting the survivors (Neyman, 1937). Viewed in this manner, the term "statistically insignificant" is rather weak – it simply indicates that zero was among the set of survivors. An insignificant coefficient does not mean that zero is the true value or even the best guess, it simply states that zero – among others – is a survivor among every other unrejected value, all compatible with the data at the 5 percent significance level. The informativeness of an insignificant result then is not the fact that zero lies in the interval, but rather, which other coefficient effect sizes survived, and which were rejected.

Consider two studies, each reporting a statistically insignificant coefficient with a p-value of approximately 0.5. Study A reports a 95 percent confidence interval of [−0.02, 0.04] on an economic outcome whose economically meaningful effects, as documented in the literature, begin at about $|\beta| = 0.1$. Study B finds the same confidence interval – using a smaller sample – to be [−0.45, 0.48]. Though the reading of the variable of interest in the regression table is almost identical – both "statistically insignificant" – their confidence intervals contain starkly different information. The results of Study A tell us that economically meaningful effects are rejected at the 5 percent level, and that any effect, if present, is too small to be of economic consequence. The Study B interval is far less informative as it suggests that effects of every magnitude – negligible, moderate, and large, in either direction – remain unrejected. Study B is unable to distinguish between any hypotheses that would be of value to the reader, and its insignificant coefficient estimate simply indicates that the study was run and no meaningful outcome could be discerned. Figure 1 depicts the contrast, together with a conventionally significant result for comparison.

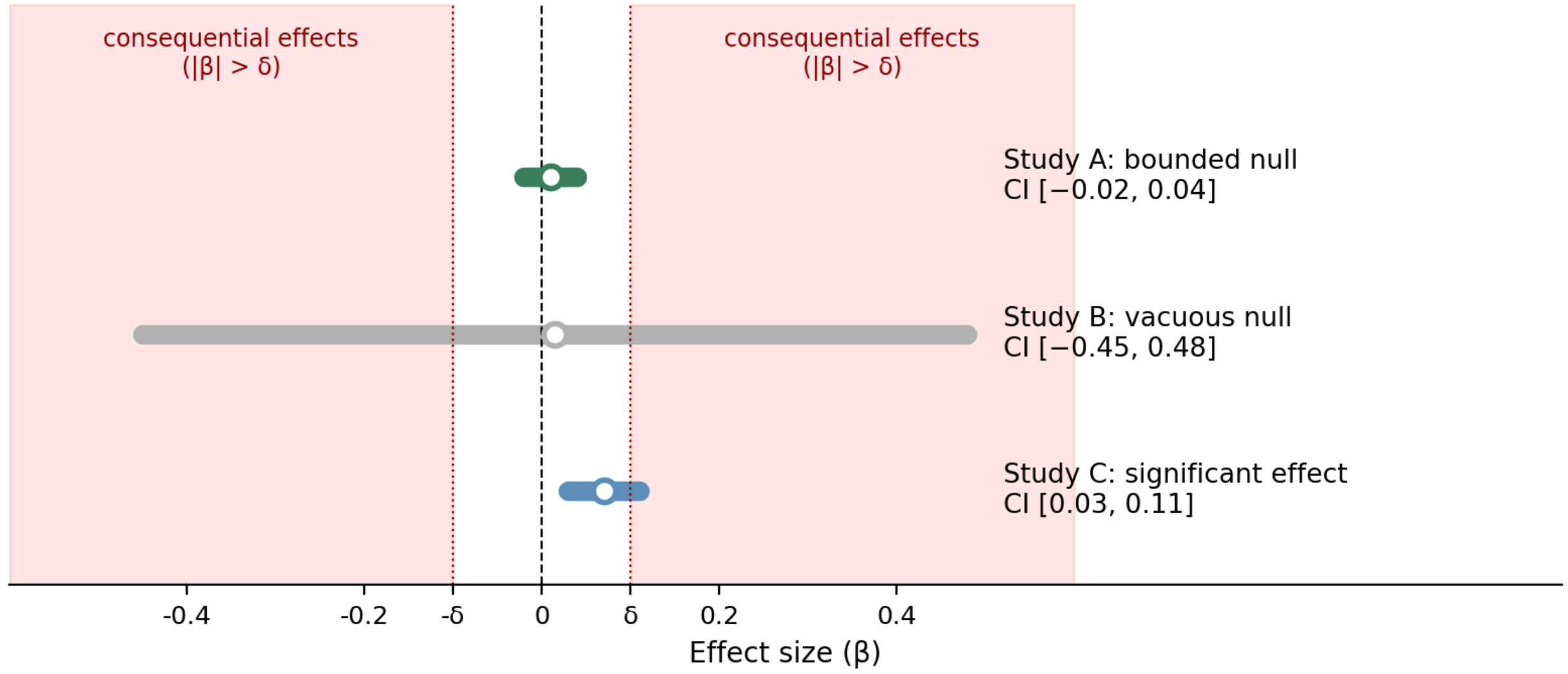

***Figure 1.*** *Three regression results. Studies A and B are both "insignificant" — both intervals contain zero — but Study A excludes all effects that matter (|β| > δ) while Study B does not. Only Study B's null is uninformative.*

I introduce the following protocol when interpreting an insignificant coefficient of a null claim. The *relevant edges* of an interval lie on the side(s) the claim addresses: both edges for a two-sided claim ("no effect"), and the relevant side for a one-sided claim ("no penalty", "no benefit").

1. What is the largest effect size at the relevant edge(s)? If no interval is provided, reconstruct this from the reported coefficient and standard error.

2. Is the largest effect size in the relevant edge(s) of economic importance? I introduce δ: the smallest effect size of consequence based on the context of the study.

3. If the answer is no, then the null claim for the relevant edge(s) is bounded; that is, the null claim ("X has no effect on Y") is warranted by the data. If the relevant edge(s) does include effect sizes of consequence, then the null claim is vacuous as the data does not warrant the assertion.

Note that a two-sided null claim can arrive at a split verdict where one edge is bounded and another is vacuous; that is, it overclaims in one direction while being bounded in the other.

None of the steps in the above protocol require data. Step 1 needs only the coefficient estimate and standard error – readily observable in all standard regression tables. Steps 2 and 3 only require a benchmark for what constitutes a meaningful effect size – discussed in the following section. A formal counterpart exists – equivalence testing, which converts the same interval-versus-δ comparison into a hypothesis test with the burden of proof reversed (Lakens, 2017) – of which the protocol above is the informal, everyday form. Kim and Robinson (2019) present these tests for economics and finance settings explicitly.

The bounded null claim is not a statement on the true value of the effect size (for example, statements like "we conclude that the effect size is zero") as the confidence interval includes a range of values that remain unrejected by the data at the requisite α. Rather, it is a statement on what is *excluded* from the set of surviving values: "Effect sizes larger than δ are rejected by the data". The vacuous null claim makes no such exclusion – consequential effect sizes and zero alike cannot be rejected – and is thus uninformative: the study lacked the precision to distinguish meaningful effects from none.

## 3. What counts as an effect that matters

Perhaps the most important step in the protocol is to determine what is a consequential economic effect size versus a negligible one. I define δ as the smallest effect size of consequence[3] – per increment of $X$ – based on the context of the study and the dependent variable. There is no universal threshold. For example, what makes 0.02 negligible or material depends on whether a change of this size would alter any decision that turns on the outcome; and is thus largely context specific. Magnitude-based interpretation has historically been critiqued for being subjective and rhetorical: if δ is simply a judgment determined by the researcher, then – as the objection goes – any classification is simply opinion dressed up as measurement.

The current convention reports a statistically insignificant coefficient and concludes that the population value is essentially zero and has no material impact on the dependent variable. However, this practice itself is making a judgment on δ, albeit a silent, arbitrary and unreported one. A conventional reading of a statistically insignificant coefficient assumes that δ lies beyond the relevant edges of the confidence interval, whatever the edge values may be. Researchers have thus unknowingly made judgements on $\delta$[4]. The proposed protocol requires the author to clearly define $\delta$, source it, and defend it; and subsequently, distinguish between bounded and vacuous null claims. A stated $\delta$ can be contested; an unstated one cannot even be seen.

[3] The "smallest effect size of interest" in the equivalence-testing literature (Lakens, 2017).

The threshold, $\delta$, lives in the real world. Whether an effect size is material depends on its impact on the decision-making process in practice, and importantly, these decisions are typically made in the raw units of the dependent variable, such as percentage points of efficiency, basis points of a spread, dollars of value. As such, $\delta$ should be stated in the raw units of the dependent variable, per a named and economically meaningful increment of the regressor – a rating band, a percentage point, a feasible intervention – since a coefficient is denominated in both variables' units. It should be anchored to effect sizes that the relevant literature and decision-makers in practice report as consequential: the magnitudes prior studies report as material, the effect sizes that would impact a policy or a managerial choice, meta-analytic syntheses if available. Mitton (2024) compiles distributions of standardized effect sizes for common dependent variables in corporate finance, which is another source from which credible thresholds can be argued.

A "no cost" null claim is bounded if the relevant edge – the largest unrejected cost – of the confidence interval is below threshold $\delta$, meaning that any impact on cost is negligible and has no bearing on the decision-making process. A "no benefit" null claim is similarly bounded when the relevant edge – largest unrejected benefit – is within the $\delta$ threshold, indicating that the beneficial effect (if any) is not material enough for the decision-maker to fund. If the outcome variable has no natural units – such as an index, a standardized score, a factor loading – then raw units will carry no meaning, and the threshold should instead be set relative to the outcome and regressors' variation – for example, the effect of a one-standard-deviation movement in the regressor, expressed in standard deviations of the outcome. And again, it is up to the researcher to determine what effect size (for example, larger than one-tenth of a standard deviation) is deemed consequential.

[4] The author is a prime example of this practice, having read the insignificant coefficients of his own doctoral thesis as zeros for sixteen years.

Where should this threshold statement be situated? I argue that it belongs in the data section, while it is defended and expounded in the discussion of results. The statement in the data section is simply one paragraph – immediately following the descriptive statistics outlining the dispersion of variables – in which the author declares what is an economically consequential coefficient size (its number, its units, and its source). The burden in the data section is light; perhaps two or three sentences connecting $\delta$ to what constitutes a material magnitude in the literature or practitioner sources, and relating this to the outcome's own dispersion just reported in the descriptive statistics. For example: "Before turning to the results, we state what magnitude we would regard as economically consequential, so that insignificant estimates are read against a standard set in advance. Efficiency changes of two percentage points or more are treated as material in this literature (…), and correspond to roughly half the typical year-to-year movement in our sample; we therefore set $\delta = 0.02$". Stating $\delta$ before the results allows the reader to familiarize themselves with the threshold before encountering the regression results in Table 5, thus comparing the estimated coefficients (and intervals) against a bar already established.

It is in the discussion section, after the reporting of the estimation results, that the verdict of null claims is determined: whether they are bounded or vacuous by comparing the relevant edges to the pre-stated $\delta$. And it is there that the author engages the threshold at whatever depth the interpretation deserves – why this $\delta$ rather than a stricter or looser one, what a skeptic applying a different standard would conclude, and how a split verdict should be read.

## 4. An illustration

I will now work through a stylized example to showcase the protocol at work. Suppose there is a panel study that relates a firm outcome – scored on a unit interval with a standard deviation of 0.16 – to a regressor variable measuring a rating on a 12-band scale, making a one band change its natural unit of measure. In this example, a movement of three or four bands in the regressor represents a large, achievable real-world change. Now suppose the panel regression results report an estimated coefficient of 0.008 with a standard error of 0.013. The author – in a conventional setting – would declare the result statistically insignificant and conclude that the rating has "no effect" on the firm outcome, a common claim that appears in the ESG and corporate governance–performance literature.

Under our protocol, the author would make a statement on the threshold, δ, in the data section: say, according to related studies, outcome changes of two percentage points or more per band are considered material. As such, $\delta = 0.02$ – a threshold that is set in the outcome's raw units. Using only the estimated coefficient and standard error, the 95 percent confidence interval is calculated to be approximately [−0.018, +0.034]. The claim that $X$ "has no effect" has two sides, so both edges of the interval are relevant; and in this case, it's a split verdict. At the lower end, the relevant edge is smaller than δ ($0.018 < 0.02$), indicating that the null cost claim is bounded – any cost incurred due to a reduction in the band of $X$ results in only a negligible effect on the firm outcome – immaterial and indistinguishable from zero. However, the upper edge of the interval exceeds δ ($0.034 > 0.02$), meaning the null benefit claim is vacuous and uninformative as the author is unable to distinguish negligible and meaningful beneficial effects.

The honest summary is therefore neither "no effect" nor "uninformative" but directional: the data warrants the claim that the ratings impose no material costs while remaining silent on any claims about the upside benefits. Simply using the stated threshold and the confidence interval, I am able to transform the "no effect" claim to two precise statements – one of which is a bounded null against harm, and the other informing that any claims about benefits are unwarranted by the data. Two features of this case should be noted: first, the threshold is defined per band rather than per numerical unit, so had the rating been rescaled, coefficient and δ would rescale together and the verdict would not move. And second, a demonstrated bound on cost, and claims of benefit left unwarranted – each informative to the literature, and both concealed by the conventional "no effect" phrase.

## 5. Implications, and the audit ahead

For authors, I propose a new reporting standard: whenever a null claim on the regressor of interest is made, it should be accompanied by its confidence interval's relevant edge in the units of the decision, such as "We can rule out effects larger than δ per [increment]". As Section 3 emphasizes, δ should be stated and sourced in advance. Importantly, if the interval cannot reject effect sizes of consequence, then this should be articulated plainly. This suggests that the experiment is unable to distinguish between zero, negligible and meaningful effects. Ideally, referees should begin asking "what is the largest effect your interval still permits, and does it matter?" when reviewing null claims.

We must overturn the norm of reading "statistical insignificance" to mean zero economic effect; rather, it tells us that zero could not be rejected along with a range of other effect sizes, and whether that range is consequential is what matters. The implications for the existing literature are somewhat unsettling. In finance, there is an unknown proportion of "no effect" claims of each kind: i) bounded nulls that were undersold as mere insignificance, and ii) vacuous null claims that were oversold as evidence of absence. In economics, Fitzgerald's (2025) audit indicates that this is a substantial share. Which claims are which is an empirical question, and an answerable one. The classification of bounded and vacuous nulls requires only the reported coefficient and standard error. As such, in ongoing work, I apply this framework systematically to null claims in the finance literature, beginning with my own research. In particular, whether the imprecise intervals that follow "controlling for endogeneity" can exclude the effects the same papers report as significant under simpler estimators. The results will be reported in a companion paper.